\documentclass{pasj00}
\usepackage{natbib}
\usepackage{lscape}
\usepackage{graphicx}
\draft

\begin{document}
\SetRunningHead{Author(s) in page-head}{Running Head}
\Received{2008/06/27}
\Accepted{2008/08/06}

\title{Catalogue of $^{12}$CO(J=1-0) and $^{13}$CO(J=1-0) Molecular Clouds in the Carina Flare Supershell}

\author{Joanne. R. \sc{Dawson}}
\email{joanne@a.phys.nagoya-u.ac.jp}

\author{Akiko {\sc Kawamura}}
\author{Norikazu \textsc{Mizuno}}
\author{Toshikazu {\sc Onishi}}
\and
\author{Yasuo {\sc Fukui}}
\affil{Department of Physics and Astrophysics, Nagoya University, Chikusa-ku, Nagoya, Japan}

\KeyWords{Catalogs -- ISM: bubbles -- ISM: clouds -- ISM: molecules} 

\maketitle

\begin{abstract}
We present a catalogue of $^{12}$CO(J=1-0) and $^{13}$CO(J=1-0) molecular clouds in the spatio-velocity range of the Carina Flare supershell, GSH 287+04-17. The data cover a region of $\sim66$ square degrees and were taken with the NANTEN 4m telescope, at spatial and velocity resolutions of $\sim2.6'$ and $0.1$ km s$^{-1}$. Decomposition of the emission results in the identification of 156 $^{12}$CO clouds and 60 $^{13}$CO clouds, for which we provide observational and physical parameters. Previous work suggests the majority of the detected mass forms part of a comoving molecular cloud complex that is physically associated with the expanding shell. The cloud internal velocity dispersions, degree of virialization and size-linewidth relations are found to be consistent with those of other Galactic samples. However, the vertical distribution is heavily skewed towards high-altitudes. The robust association of high-$z$ molecular clouds with a known supershell provides some observational backing for the theory that expanding shells contribute to the support of a high-altitude molecular layer. 
\end{abstract}

\section{Introduction}

The mechanical feedback from OB clusters has a major impact on the structure and evolution of the interstellar medium. Supershell shocks are known to drive correlated episodes of triggered star formation -- either through the compression of existing molecular material to an unstable state \citep[e.g.][]{boss95, vanhala98}, or through the manufacture and subsequent collapse of parent molecular clouds from the accumulated gas \citep{bergin04, hartmann01, mccray87, koyama00}. Observational studies of molecular supershells have the power provide constraints on theories of triggered cloud and star formation, and to accurately pinpoint locations at which such triggering may be occurring. Yet, with a handful of exceptions \citep{fukui99, jung96, matsunaga01, moriguchi02, normandeau96, rizzo98,  yamaguchi99, yamaguchi01b}, supershells with a significant molecular component remain relatively understudied.\par

The CO molecule has been widely and successfully used as a tracer of the molecular ISM, and has formed the basis of many large-scale surveys of the Galactic molecular disk \citep[e.g. ][]{dame87, heyer98, mizuno04, sanders86}. Uniform, wide-coverage CO datasets are indispensable for the information they provide on the distribution and physical properties of the molecular ISM, and the decomposition of such datasets into isolated areas of emission enables the statistical analysis of large populations of clouds in a variety of environments \citep[e.g.][]{fukui08, heyer01, kawamura98, solomon87, yonekura97}. 

This paper presents a catalogue of CO molecular clouds in the region of the `Carina Flare' supershell 
(GSH 287+04-17). The Carina Flare is an H{\sc i}-H$_2$ supershell located in the near side of the Carina Arm at $D=2.6\pm0.4$ kpc. It was first discovered in $^{12}$CO(J=1-0) with the NANTEN 4m telescope by \citet{fukui99}, who reported the discovery of an extensive, expanding complex of molecular clouds reaching as high as $z\sim450$ pc above the Galactic Plane. Comparisons with H{\sc i} 21 cm survey data confirmed that the majority of the mass in this molecular complex -- as well as much additional lower-latitude material -- forms co-moving parts of a gently expanding atomic supershell \citep{dawson08}. This supershell extends from $b\sim0.5^{\circ}$ to $b\sim10^{\circ}$, covering a total projected area of $\sim70$ square degrees, with a main body $360\times260$ pc in size, and a capped, high-latitude extension. The total associated molecular mass is estimated as $M(\mathrm{H}_2)=2.0\pm0.6\times10^5 M_{\solar}$, which amounts to $\sim20\%$ of the total neutral gas mass, and this molecular component is associated with a rich variety of substructural features in the atomic shell. Moreover, several molecular clouds show evidence for active star formation, suggesting triggering may be occurring \citep{dawson07, fukui99}. 

Previous work has discussed the macroscopic distribution and properties of the Carina Flare molecular ISM. However, detailed data on individual clouds, as well as the statistical properties of the sample as a whole, have not previously been published. Here we present the first comprehensive catalogue of the cloud population, intended both as a complement to past work, and as a resource upon which future studies of the region may draw. The present study includes observations in both $^{12}$CO(J=1-0) and $^{13}$CO(J=1-0). The $^{12}$CO line, which is generally assumed to be optically thick, is an excellent tracer of the overall molecular gas distribution, down to diffuse, low-luminosity envelopes. This is complemented by the $^{13}$CO line, which is optically thin up to H$_2$ column densities of $N(\mathrm{H}_2)\sim10^{22}$ cm$^{-2}$, and correlates well with denser material and star forming regions.  

As well as the obvious interest that comes from investigating the molecular component of a known supershell, the present dataset also has several advantages that make it particularly attractive for this kind of large-scale cloud catalogue. Firstly, the association of the molecular clouds with a single object means that their distances are both uniform and well-determined, reducing what would otherwise be a substantial source of uncertainty in the determined cloud properties. Secondly, the majority of the complex is at high enough latitudes that line-of-sight confusion is minimal. This means that emission may be assigned to individual clouds without the need to resort to artificially high intensity thresholds, and leads to the inclusion of low-luminosity cloud envelopes that may be missed by other studies. In addition, the present dataset is both sensitive and relatively well resolved, with rms noise levels per 0.1 km s$^{-1}$ channel of $< 0.5$ K in $^{12}$CO(J=1-0) and $< 0.2$ K in $^{13}$CO(J=1-0), and telescope half power beam widths of $\sim2.6'$ and $\sim2.7'$ respectively. 

The paper is organized as follows. Section \ref{observations} describes the instrumentation and observing strategy. Section \ref{cat} presents the $^{12}$CO and $^{13}$CO cloud catalogue tables, and   also describes the cloud decomposition method, the derivation of cloud properties and the completeness and coverage of the catalogue. Section \ref{results} examines some of the statistical properties of the cloud sample as a whole, and section \ref{discussion} briefly discusses the vertical distribution of the molecular gas, as well as some of the issues surrounding mass estimation. We summarize in section \ref{summary}.\\\par

\section{Observations}
\label{observations}

Observations of the $^{12}$CO(J=1-0) and $^{13}$CO(J=1-0) transitions at 115.3 GHz and 110.2 GHz were made with the NANTEN 4m telescope at Las Campanas Observatory in Chile. The 4 K cryogenically cooled SIS mixer receiver provided a typical system noise temperature of $\sim$250 K at 115 GHz, and $\sim130$ K at 110 GHz (in a single side band), including the atmosphere towards the zenith \citep{ogawa90}. Intensity calibration was made by the chopper wheel method. A 2048 channel acousto-optical spectrograph provided a total bandwidth of 40 MHz and an effective spectral resolution of 40 kHz, corresponding to a velocity coverage and resolution of 100 km s$^{-1}$ and 0.1 km s$^{-1}$, respectively. Orion KL ($\alpha_{B1950}=\timeform{5h32m47.5s}, \delta_{B1950}\timeform{-5D24'21''}$) was observed as a standard calibrator source, with absolute antenna temperatures of $T_R^*=65$ K and 10 K assumed for the $^{12}$CO and $^{13}$CO lines. The telescope half power beam width was $\sim2.6'$ at 115 GHz and $\sim2.7'$ at 110 GHz. \par

The data presented in this catalogue were taken in two observing runs covering the periods April to November 1998 and November 2001 to May 2002. The former were originally published by \citet{fukui99} and the latter formed part of the PhD thesis of \citet{matsunagaphd}. Both datasets were also included by \citet{dawson08} in their study of the atomic and molecular components of the supershell ISM. 

\begin{figure}
  \begin{center}
   \FigureFile(160mm,160mm){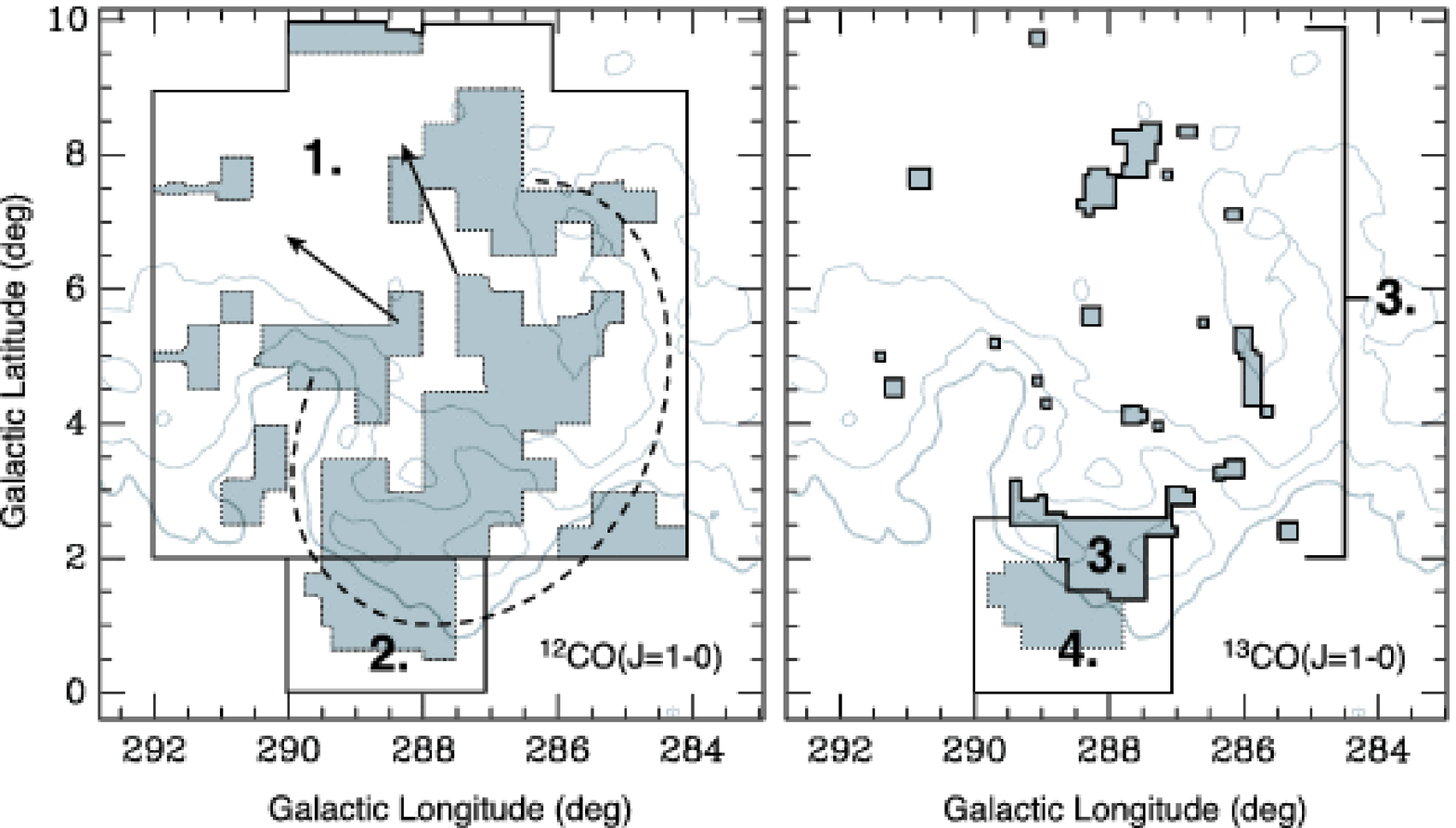}
  \end{center}
  \caption{Observed regions. Regions observed in $^{12}$CO(J=1-0) (left) and $^{13}$CO(J=1-0) (right) are enclosed in solid lines. Grey areas are covered at a grid spacing of $2'$ and all other regions at $4'$. Light grey contours are low-resolution Parkes telescope H{\sc i} Southern Galactic Plane Survey (SGPS) data \citep{mcclg05} integrated over the velocity range $-36 < v_{lsr} < -5$km s$^{-1}$ (contour levels are 880, 1150 and 1420 K km s$^{-1}$). The approximate outline of the shell, measured at its central velocity ($\approx-17$ km s$^{-1}$), is taken from \citet{dawson08} and shown here by the thick dashed line. Arrows indicate the direction in which the shell is thought to be blowing out of the Galactic Plane.}
  \label{obsregions}
\end{figure}

Details of the observing strategies, spatial coverage and data sensitivity are given in table \ref{obstable} and figure \ref{obsregions}. Pointing centers for all regions were arranged in a square grid with separations of either $2'$ or $4'$ between adjacent pointings. Region 1 was initially surveyed in $^{12}$CO(J=1-0) at a spacing of $8'$, and this preliminary data used as a guide for targeted $2'$ observations. $^{13}$CO(J=1-0) pointings were then made at $2'$ spacing towards clouds for which the peak $^{12}$CO(J=1-0) antenna temperature exceeded $\sim4$ K (region 3). The remainder of region 1 was later covered in $^{12}$CO(J=1-0) at a spacing of $4'$, to attempt to pick up clouds missed by the initial $8'$ pass. Regions 2 and 4 were initially covered at $4'$ spacing in both $^{12}$CO(J=1-0) and $^{13}$CO(J=1-0),  with targeted $2'$ follow-up observations made towards the giant molecular cloud (GMC) G288.5+1.5 \citep{matsunagaphd}. The r.m.s noise per channel in $T_R^*$ is $\lesssim0.5$ K at 115 GHz and $\lesssim0.2$ K at 110 GHz.\\\par.

\begin{table}
\begin{center}
  \caption{test caption}
  \label{obstable}
  \begin{tabular}{lllll}
  \hline              
  & $^{12}$CO(J=1-0) & & $^{13}$CO(J=1-0) &\\ 
   \hline
   Region number\footnotemark[1]{} & 1 & 2 & 3 & 4 \\
   Pointing grid & $2'$, $4'$ & $2'$, $4'$ & $2'$ & $2'$, $4'$\\
   Observing strategy\footnotemark[2]{} & pos. sw. & pos. sw. & freq. sw. & pos. sw.\\
   RMS noise per channel & $\sim0.5$ K & $\sim0.3$ K & $\sim0.2$ K & $\sim0.1$ K \\
   Observing period & $04/98 - 04/99$ & $11/01 - 02/02$ & $11/98$ & $02/02 - 05/02$\\
   Publications\footnotemark[3]{} & F99 & M02 & F99 & M02 \\
   \hline
   \end{tabular}
\footnotetext[1]{Refers to figure \ref{obsregions}}\\
\footnotetext[2]{Whether observations were made by position or frequency switching}
\footnotetext[3]{F99: \citet{fukui99}. M02: \citet{matsunagaphd}}
\vspace{0.7cm}
\end{center}
\end{table}

\section{The Catalogue}
\label{cat}

\subsection{Spatial and Velocity Coverage}

The present catalogue covers an area of $\sim66$ square degrees, encompassing $\gtrsim90\%$ of the Carina Flare supershell's projected area \citep{dawson08}. The only significant omission is a small section on the lower right-hand rim, for which high quality NANTEN data is not available. The catalogue velocity coverage is chosen to fully encompass the extent of the supershell in $v_{lsr}$, and includes all CO clouds whose peak velocities fall in the range $-40 < v_{lsr} < 0$ km s$^{-1}$. Emission in this range accounts for $\sim85\%$ of the total velocity integrated intensity between -50 and +50 km s$^{-1}$ in the catalogue spatial area. Of the total emission in the catalogue spatio-velocity cube, $\gtrsim90\%$ is contained in regions observed at $2'$ grid spacing. This is true for both the $^{12}$CO and $^{13}$CO data sets.\\\par

\subsection{Cloud Definition}

We consider as positive detections all points at which emission is detected above the $3\sigma$ noise level in integrated intensity, where $\sigma$ is the standard deviation in integrated intensity over a typical linewidth in emission-free sections of the datacube, computed for the noisiest regions of the two datasets. For the $^{12}$CO(J=1-0) dataset $3\sigma\sim1.5$ K km s$^{-1}$ and for $^{13}$CO(J=1-0) the value is $\sim0.45$ K km s$^{-1}$.  

\begin{landscape}

\begin{figure}
  \begin{center}
    \FigureFile(210mm,210mm){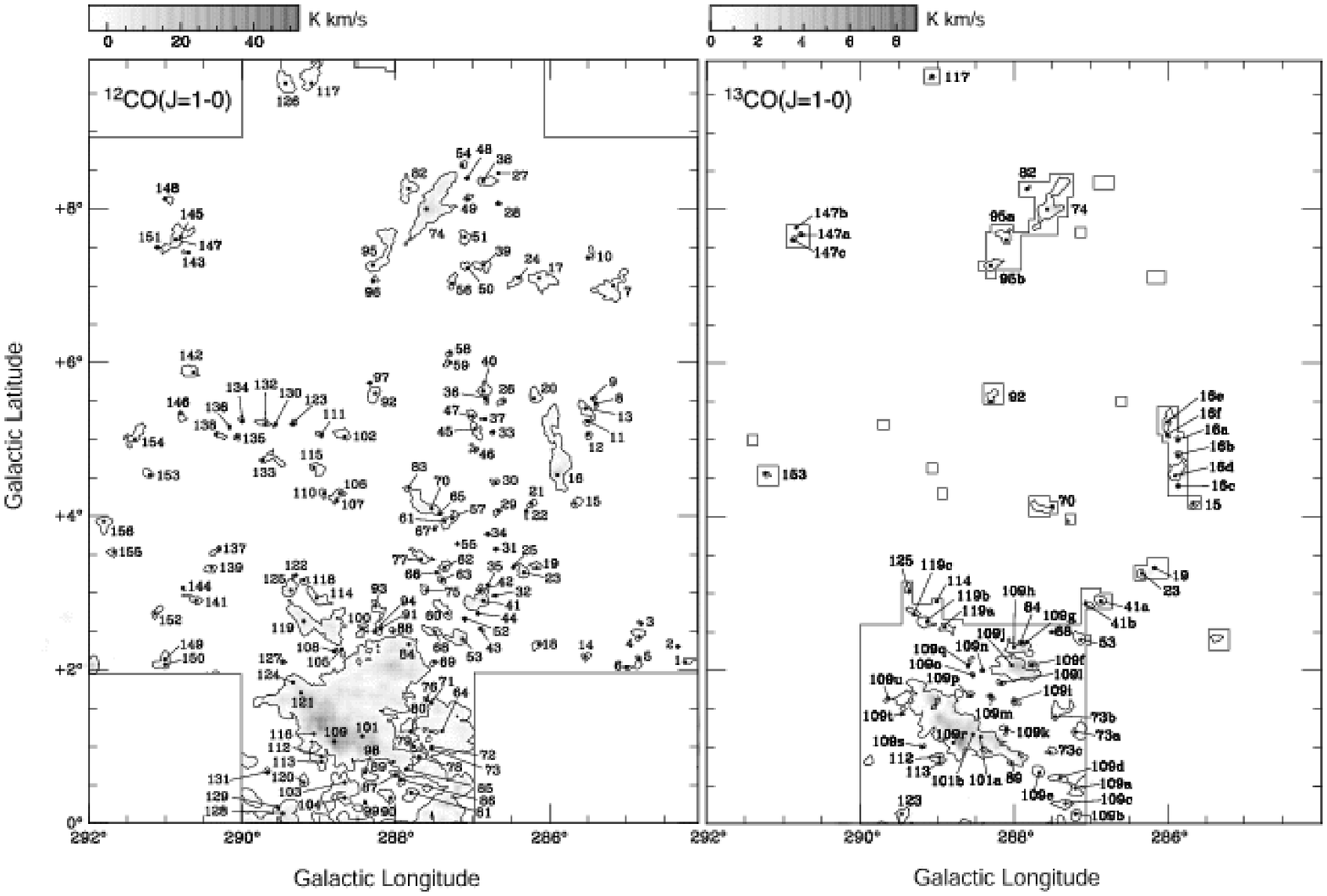}
  \end{center}
  \caption{Finding chart for the Carina Flare $^{12}$CO(J=1-0) (left) and $^{13}$CO(J=1-0) (right) molecular clouds. The emission is integrated over the velocity range $-40 < v_{lsr} < 0$ km s$^{-1}$, and contours are drawn at the $3\sigma$ detection limit, corresponding to 1.5 K km $s^{-1}$ for $^{12}$CO and 0.45 K km $s^{-1}$ for $^{13}$CO. Cloud numbers refer to those in the catalogue tables, and black dots mark the positions of peak intensity for each cloud.}
  \label{findingboth}
\end{figure}

\end{landscape}

We assign detected pixels into molecular clouds, where a cloud is defined as a discrete region of detected emission that is contiguous in $(l,b,v_{lsr})$ space. For emission to be considered contiguous in position it is required to occupy adjacent or diagonally adjacent grid points. For emission to be considered contiguous in velocity, Gaussian profiles fitted to each spectral line must overlap before reaching the rms noise temperature. In practice this usually corresponds to a velocity separation of $\lesssim5$ km s$^{-1}$. For two positionally overlapping clouds to be considered as separate entities, their velocities must be well separated across all observed points. Clouds whose emission extends outside the catalogue $(l,b,v_{lsr})$ space are excluded, and we also reject single points with no surrounding emission, which are not considered robust detections. This results in the final decomposition of the emission into 156 $^{12}$CO clouds and 60 $^{13}$CO clouds. Finding charts for these are given in figure \ref{findingboth}.\\\par

\subsection{Observational Properties of the CO Clouds}

For velocity channel maps and additional plots of the distribution of the CO clouds, readers are referred to \citet{fukui99} and \citet{dawson08}.\par 
The observed properties of the $^{12}$CO and $^{13}$CO clouds are summarized in tables \ref{obsproptable12} and \ref{obsproptable13}. Column (1) gives cloud numbers (corresponding to those in figure \ref{findingboth}), and column (2) gives $^{12}$CO cloud names. $^{13}$CO clouds are not assigned unique names, but are instead assigned numbers according to their parent $^{12}$CO clouds. Column (3) gives the angular separation between adjacent pointings. Columns (4-7) give the positions of peak integrated intensity for each cloud in both galactic and equatorial coordinates. Columns (8-11) give basic line properties at the position of peak integrated intensity; the absolute peak antenna temperature, $T_R^*$, the integrated intensity, $W_{\mathrm{CO}}$, the FWHM velocity dispersion, $\Delta v$, and the center velocity with respect to the local standard of rest, $v_{lsr}$. All properties are derived from gaussian fits to the line profiles. In the vast majority of cases profiles are well fitted by a single gaussian and the relevant parameters are readily recovered. In the rare cases where multiple components are present $v_{lsr}$ is given by the intensity-weighted mean velocity, and $\Delta v$ is defined as the intensity-weighted standard deviation $\sigma_v$ multiplied by the factor $\sqrt{8ln(2)}$ \citep[see also][]{rosolowsky06}. Columns (12-13) give the internal velocity dispersion and average center velocity of each cloud as a whole, $\Delta v_{cld}$ and $v_{lsr, cld}$. These are derived as above from gaussian fits to the arithmetic mean of all spectra within a given cloud. In all cases where multiple components are fitted we also explicitly include parameters for each component in the catalogue tables (given in round brackets). It should be noted that the calculation of these properties from the fitted gaussian functions results in typical deviations of only $\lesssim2\%$ compared to values calculated directly from the data.\\\par 

\subsection{Derived Properties of the CO Clouds}
\label{derivedprops}

\subsubsection{$^{12}$CO Cloud Properties}
\label{derivedprops12co}

Derived properties of the $^{12}$CO clouds are given in table \ref{physproptable12}. A distance of 2.6 kpc is assumed throughout. Column (7) gives the projected area of the cloud in parsecs, $A_{cld}$. This is calculated by summing the areas of all pixels contained within a given cloud and subtracting an approximate correction factor to account for the spreading effect of the $2.6'$ beam. The correction factor is computed by deconvolving a circular, gaussian intensity distribution of FWHM $=\sqrt{A_{cld}/\pi}$ with the telescope beam profile.
Column (8) gives the CO luminosity, $L_{\mathrm{CO}}$, defined as the sum of the velocity-integrated intensities over the entire projected area of the cloud:
\vspace{0.5cm}
\begin{equation}
L_{\mathrm{CO}} = \sum{~[D(\Delta\Omega_i)]^2~W_{\mathrm{CO}, i}} 
\end{equation}
\vspace{0.5cm}
\noindent where $D$ is the distance to the molecular cloud, $\Delta\Omega_i$ is the angular size in radians of pixel $i$ and $W_{\mathrm{CO},i}$ is the integrated intensity of pixel $i$. 
Column (9) gives the CO luminosity mass of the cloud, $M_{\mathrm{CO}}$, estimated from the empirical relationship between $^{12}$CO(J=1-0) integrated intensity and molecular hydrogen column density; $N(\mathrm{H}_2)=XW_{\mathrm{CO}}$. Thus,
\vspace{0.5cm}
\begin{equation}
M_{\mathrm{CO}} = X\mu m_HL_{\mathrm{CO}}.
\end{equation}
\vspace{0.5cm}
\noindent Here, in keeping with F99, we have adopted the value of the conversion factor derived by Hunter et al (1997), $X = 1.56\times10^{20}$ K$^{-1}$ km/s cm$^{-2}$. The molecular weight, $\mu$, is taken as 2.7 to account for the presence of helium and heavier elements.
We also calculate virial masses for the clouds, given in column (10). We adopt a simplified formulation that considers only gravitational and internal pressure, and ignores the effect of external pressure and magnetic fields. Assuming a spherically symmetric cloud with a $1/r$ density distribution,
\vspace{0.5cm}
\begin{equation}
M_{vir}/M_{\solar} = 190~R (\Delta v_{cld})^2 
\end{equation}
\vspace{0.5cm}
\noindent where $R$ is the effective radius in parsecs, given by $R = \sqrt{A_{cld}/\pi}$, and $\Delta v_{cld}$ is the cloud velocity dispersion in km s$^{-1}$. \\\par

\subsubsection{$^{13}$CO Cloud Properties}

Derived properties of the $^{13}$CO clouds are given in table \ref{physproptable13}.
Column (7) gives the projected area of the cloud, $A_{cld}$, defined as above.
Column (8) gives the optical depth, $\tau(^{13}\mathrm{CO})$ at the peak of each $^{13}$CO cloud, where $\tau(^{13}\mathrm{CO})$ is computed directly from the absolute antenna temperature as follows:
\vspace{0.5cm}
\begin{equation}
\tau(^{13}\mathrm{CO})= -\ln \left[1-\frac{T_R^*(^{13}\mathrm{CO})}{5.29\left(\left\{\exp(5.29/T_{ex})-1\right\}^{-1}-0.164\right)}\right].
\end{equation}
\vspace{0.5cm}
Column (9) gives the peak molecular hydrogen column density, $N(\mathrm{H}_2)$. This is derived from the $^{13}$CO column density, assuming an H$_2$ to $^{13}$CO abundance ratio of $5\times10^5$ to 1 \citep{dickman78}. The $^{13}$CO column density is given by the following expression, under the assumption that the J=1 and J=0 rotational levels of the molecule are in local thermodynamic equilibrium (LTE):
\vspace{0.5cm}
\begin{equation}
N(^{13}\mathrm{CO})= 2.42\times10^{14}\left[\frac{T_{ex}~\tau(^{13}\mathrm{CO})~\Delta v}{1-\exp(-5.29/T_{ex})}\right],
\end{equation}
\vspace{0.5cm}
\noindent where the excitation temperature, $T_{ex}$, is assumed to be 15 K for all clouds. 

Column (10) shows the cloud LTE mass, computed by summing H$_2$ column densities across all pixels in the cloud. Thus,
\vspace{0.5cm}
\begin{equation}
M_{\mathrm{LTE}} = \mu m_H\sum{~[D(\delta\Omega_i)]^2~N(\mathrm{H}_{2, i})} 
\end{equation}
\vspace{0.5cm}
\noindent where $N(\mathrm{H}_{2, i})$ is the column density of pixel $i$, and all other quantities are defined as above.
Column (11) gives the cloud virial mass, calculated in the same way as for $^{12}$CO (see \ref{derivedprops12co}).\\\par

\subsection{Completeness of the Catalogue}

The condition that emission must be detected above the $3\sigma$ noise level in integrated intensity results in a H$_2$ column density detection limit of $N(\mathrm{H}_2)\gtrsim2.5\times10^{20}$ cm$^{-2}$ for both $^{12}$CO and $^{13}$CO clouds. For $^{12}$CO this figure arises directly from the linear relationship between $W_\mathrm{CO}$ and $N(\mathrm{H}_2)$ . For $^{13}$CO clouds, where $N(\mathrm{H}_2)$ depends non-linearly on antenna temperature and linewidth, we nevertheless find a similar cutoff in detectable column density.\par
 
The variable grid spacing and the requirement that a cloud be composed of at least two adjacent spatial points dictates the completeness limit of the catalogue. For $2'$ pixels, although the sampling is not Nyquist, there is a reasonable overlap between the $2.7'$ beam at adjacent pointings, and the majority of detectable emission is picked up. Two adjacent $2'$ pixels at the column density detection limit correspond to a molecular mass of $\sim25 M_{\solar}$. However, this is not the catalogue completeness limit, since more massive clouds with small angular extents may be missed. Examination of the data indicates that isolated single-point detections are rare, and typically correspond to masses of $\lesssim30 M_{\solar}$. Considering these factors, and the slight undersampling of the data, we adopt a very conservative completeness limit of $\sim60M_{\solar}$ for the $2'$ regions of the catalogue.\par 

The $4'$ regions, on the other hand, are significantly undersampled. Resampling $2'$ regions at $4'$ recovers all clouds with masses of $\gtrsim 200M_{\solar}$, indicating that the catalogue as a whole is fully complete to around this limit.\\\par

\section{Results}
\label{results}

\subsection{Velocity Dispersion}

\begin{figure}
  \begin{center}
    \FigureFile(100mm,100mm){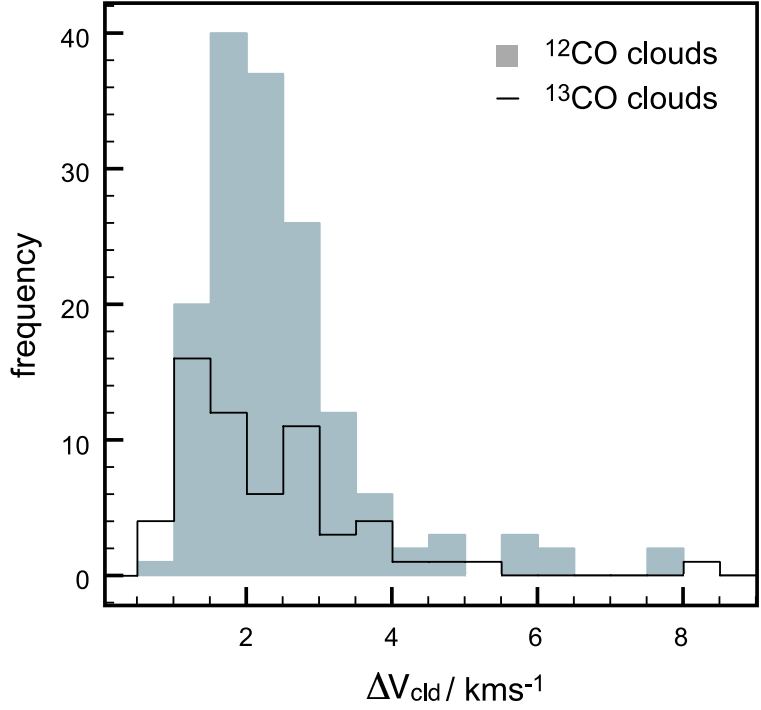}
  \end{center}
  \caption{Histograms of cloud velocity dispersion, $\Delta v_{cld}$ for $^{12}$CO clouds (grey) and $^{13}$CO clouds (solid line).}
  \label{dvhist}
\end{figure}

Figure \ref{dvhist} shows a histogram of the $^{12}$CO and $^{13}$CO cloud velocity dispersions, $\Delta v_{cld}$. The means of the two distributions are 2.5 kms$^{-1}$ and 2.3 kms$^{-1}$ for $^{12}$CO and $^{13}$CO respectively, with standard deviations of 1.2 kms$^{-1}$ and 1.3 kms$^{-1}$. Such values are typical of galactic molecular clouds in this size range \citep[e.g.][]{fujishitamast, heyer01, kawamura98, yonekura97, solomon87}, and the population as a whole shows no evidence of a systematic enhancement in $\Delta v_{cld}$ that might indicate shocked or highly disrupted molecular gas. Similarly, with one possible exception, we also find no evidence of unusually broad components in individual spectra. This exception is the GMC G288.5+1.5, which shows a complex velocity structure and exhibits velocity dispersions of $\sim10$ km s$^{-1}$ in some localized regions. This is consistent with the suggestion that this cloud lies along the interface between the Carina Flare and the Carina OB2 supershell \citep{rizzo98} and is interacting with both objects \citep{dawson08}.\\\par

\subsection{Mass Spectra}

A truncated power law of the form

\begin{equation}
N(>M) = k M^{\gamma+1} - N_0
\end{equation}

\noindent	is fitted to the cumulative mass distribution of the $^{12}$CO and $^{13}$CO clouds using the maximum likelihood method of \citet{crawford70}. The parameter $N_0$ is a measure of the number of clouds at the high mass end of the spectrum; specifically the number with $M>2^{1/(\gamma+1)M_0}$, where $M_0$ is the highest mass in the distribution, given by $M_0=(N_0/k)^{1/\gamma+1}$ \citep{rosolowsky05}. Because the completeness limit for the dataset as a whole is dragged up by undersampling in the 4$'$ regions, we consider only clouds sampled with a 2$'$ grid spacing. This results in a reduction of the cloud populations by $\sim15\%$, but allows us to achieve the maximum completeness limit of $\sim60M_\odot$ in both $^{12}$CO $M_{\mathrm{CO}}$ and $^{13}$CO $M_{\mathrm{LTE}}$.\par

\begin{figure}
  \begin{center}
    \FigureFile(165mm,165mm){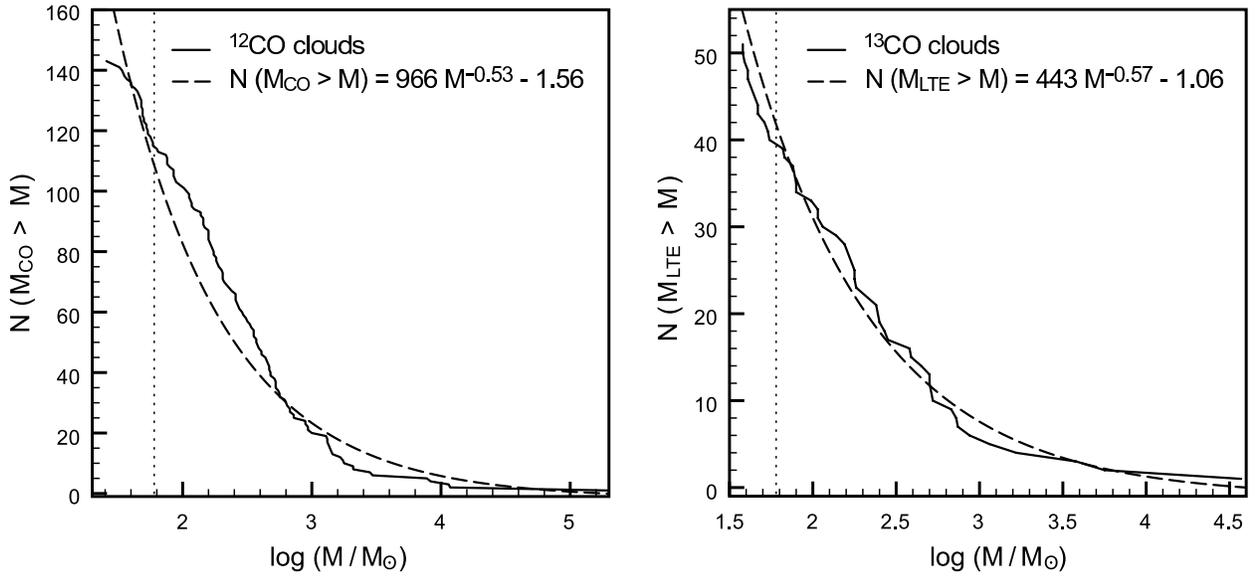}
  \end{center}
  \caption{Cumulative mass spectra of $^{12}$CO clouds (left) and $^{13}$CO clouds (right) observed at $2'$ grid spacing. Best fit power laws above the completeness limit of $\sim60M_\odot$ are shown by dashed lines.}  
  \label{mspectfig}
\end{figure}

The $^{12}$CO cloud distribution above $\sim60M_\odot$ is best fitted by a power law of $\gamma = -1.53\pm0.06$, and the $^{13}$CO by $\gamma = -1.57\pm0.11$, where the errors are the statistical uncertainties on the fits, and we make no attempt to account for uncertainties in the masses themselves. The mass spectra and their fits are plotted in figure \ref{mspectfig}. These results are typical of the inner and local galaxy, in which values of $\gamma$ between -1.4 and -1.8 are recorded for both $^{12}$CO and $^{13}$CO clouds across a wide range of masses \citep[see e.g.][and references within]{kawamura98, kawamura99, yamaguchi99, rosolowsky05}. The parameter $N_0$ is close to unity for both fits, indicating that there is no significant truncation of the functions. \par

The $^{12}$CO cloud sample departs notably from a true power law. It shows an excess of smaller clouds and a lack of larger ones in comparison to the constant-index fit; equivalent to saying that the low mass end of the spectrum would be better fitted by a shallower index and the high mass end by a steeper one. This behavior is unlikely to arise from an incorrect completeness limit -- the $60M_\odot$ estimate is a conservative one. It is therefore likely that the distribution is a real property of the cloud population.\\\par

\subsection{Virial Analysis}

The comparison between luminosity-based mass estimates and the virial mass is frequently used as a method of investigating the gravitational stability of molecular clouds. Defined as in section \ref{derivedprops12co}, the virial mass represents the mass required for a spherical cloud with a $1/r$ density distribution to be just self-gravitating against its own internal velocity dispersion, in the absence of external pressure or magnetic fields. In this analysis a cloud is described as `virialized' when its luminosity-based mass equals its virial mass, although it must be stressed that this condition does not necessarily imply a truly bound object.\par 

\begin{figure}
  \begin{center}
    \FigureFile(165mm,165mm){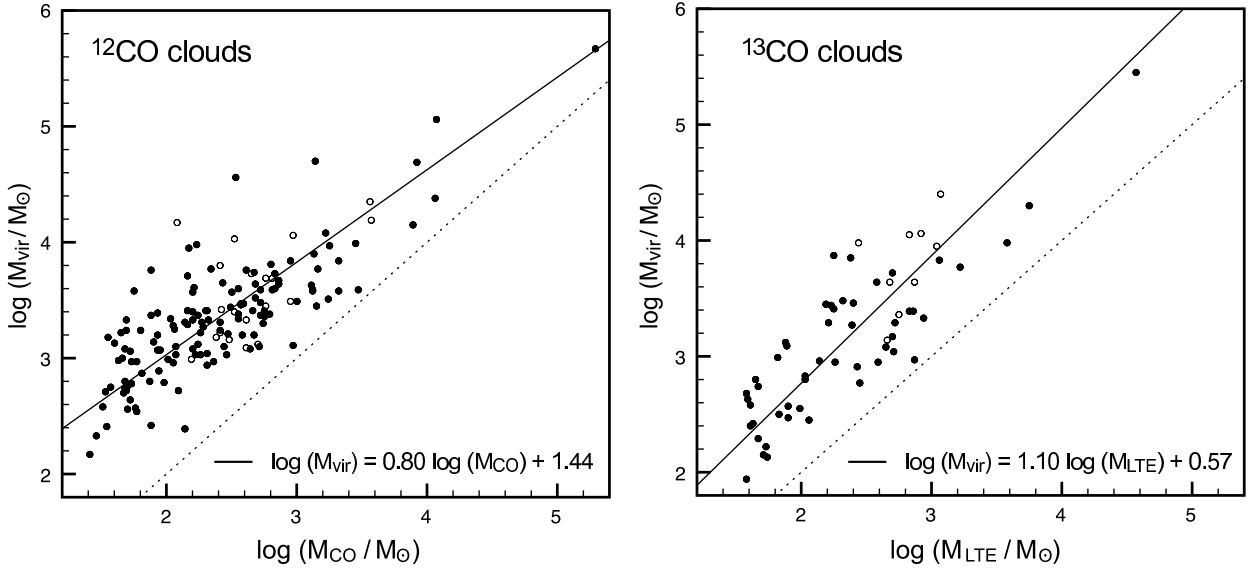}
  \end{center}
  \caption{Luminosity-based mass vs virial mass for $^{12}$CO clouds (left) and $^{13}$CO clouds (right). Filled circles mark clouds observed at $2' $ grid spacing and open circles mark those observed at $4'$ grid spacing. Solid lines are robust bisector fits to the data points. Dotted lines show $M_{\mathrm{CO}} = M_{vir}$ in the case of $^{12}$CO, and $M_{\mathrm{LTE}} = M_{vir}$ in the case of $^{13}$CO.}
  \label{virialfig}
\end{figure}

$M_{\mathrm{CO}}$ and $M_{\mathrm{LTE}}$ are plotted against $M_{vir}$ in figure \ref{virialfig}. Robust bisector fits to the distributions produce $\log(M_{\mathrm{CO}}) = 0.80 \log(M_{vir}) + 1.44$ for the $^{12}$CO clouds, and $\log(M_{\mathrm{LTE}}) = 1.10 \log(M_{vir}) + 0.57$ for the ${13}$CO clouds. The Spearman's rank correlation coefficients for the samples are 0.73 and 0.80, respectively. None of the clouds in the sample are virialized. The mean factors $M_{vir}/M_{\mathrm{CO}}$ and $M_{vir}/M_{\mathrm{LTE}}$ are 14 and 9 respectively. \par

This analysis is simplistic. Nevertheless it is a useful tool for characterizing and comparing cloud populations, and has been applied to several large-scale studies of Galactic molecular clouds. For a constant X-factor of $\sim2\times10^{20}$ K$^{-1}$ km s$^{-1}$ cm$^{-2}$ (in the derivation of $M_{\mathrm{CO}}$), virialized clouds typically begin to appear at $M_{\mathrm{CO}}\sim10^{3-4}M_{\solar}$, with an increasing trend towards virialization seen at higher masses \citep{fujishitamast, heyer01, maloney90, solomon87}. By $M \gtrsim 10^5 M_{\solar}$ most clouds are seen to be virialized, although the scatter in the distribution remains large. The $^{13}$CO(J=1-0) cloud populations of \citet{kawamura98} and \citet{yonekura97} show similar behavior.\par

With this in mind, it is notable that none of the present clouds are virialized under the current forumlation, and that the distribution shows only a weak trend (in the case of $^{12}$CO) or no trend (in the case of $^{13}$CO) towards virialization at higher masses. However, this result must be interpreted with caution. The dynamic range of the catalogue clouds is small and the number of high mass objects is too low to claim statistical significance. Moreover, the apparent difference to other populations is not extreme enough that a systematic error in the distance, combined with an underestimation of the X-factor with respect to other studies, could be ruled out as the cause. We also note that our cloud definition method may include low-luminosity cloud envelopes missed by other surveys. This would have the effect of increasing cloud radius and velocity dispersion, and hence $M_{vir}$, with less effect on luminosity-based masses, but would also tend to group together clouds that other methods might count separately. A preliminary reanalysis of a small portion of the data at a higher cut-off level indicates that such effects are important, but their effect on the overall distribution is not clear, and we defer a full study for later work.\\\par

\subsection{Size-Linewidth Relation} 

A power law relationship between molecular cloud size and velocity dispersion was first noted by \citet{larson81}, using multi-line data from the literature, and has since been investigated in some depth. It has been observed over a wide range of scales and densities, from molecular cores of $R\sim0.1$pc \citep[see][and discussion within]{goodman98}, to extended molecular clouds with $R\lesssim100$pc \citep[e.g.][]{solomon87}.\par

Compared to multi-line tracers, the density range probed by $^{12}$CO(J=1-0) and $^{13}$CO(J=1-0) is limited, and a size-linewidth correlation is not readily observed across all scales. Large-scale, multi-cloud $^{12}$CO(J=1-0) surveys focusing mainly on GMCs have tended to verify the relation \citep{solomon87, scoville87}. However, recent studies extending to smaller clouds have noted an increased scatter in the distribution, and a weakening \citep{fujishitamast} or apparent breakdown \citep{heyer01} of the power law relationship for cloud radii of a few parsecs and below. $^{13}$CO(J=1-0) studies have typically found similar behavior, with poor correlations observed in cloud populations with $R\lesssim6$pc \citep[see][and references within]{yonekura97}.\par

\begin{figure}
  \begin{center}
    \FigureFile(165mm,165mm){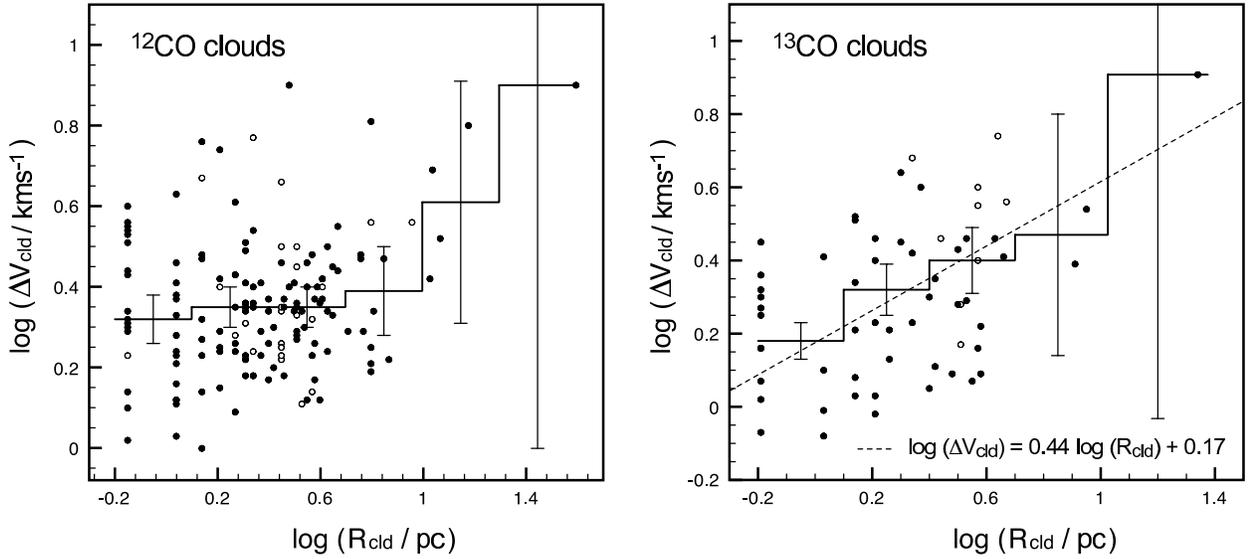}
  \end{center}
  \caption{Cloud radius vs cloud velocity dispersion for $^{12}$CO clouds (left) and $^{13}$CO clouds (right). Filled circles mark clouds observed at $2'$ grid spacing and open circles mark those observed at $4'$ grid spacing. The solid line gives the mean value of $\Delta v_{cld}$ in logarithmic bins. Error bars show the $[dN(\Delta V)]^{1/2}$ statistical errors. The dashed line is a least squares fit to the $^{13}$CO cloud sample.}
  \label{sizelwfig}
\end{figure}

In keeping with this trend, the catalogue clouds show a large scatter, and the correlation between size and linewidth is relatively weak, particularly for the $^{12}$CO clouds. Figure \ref{sizelwfig} shows plots of cloud radius, $R_{cld}$ against the cloud velocity dispersion $\Delta v_{cld}$. The mean value of $\Delta v_{cld}$ in logarithmic bins of $R_{cld}$ is also shown. For the $^{12}$CO clouds the Spearman's rank correlation coefficient is only 0.15, and there appears to be little obvious correlation below $R\sim10$ pc. The sample does show some evidence of the power-law relation appearing at larger radii, but the number of large clouds remains too small to draw statistically significant conclusions. The $^{13}$CO clouds show a somewhat tighter relation, with a Spearman's rank coefficient of 0.40, and a least squares fit produces the power law relation $\Delta v_{cld} = 1.48~(R_{cld})^{0.44}$. The difference in the behavior of the two cloud populations may reflect the different range of density regimes probed by the optically thick $^{12}$CO line and the optically thin $^{13}$CO line.
\\\par

\section{Discussion}
\label{discussion}

\subsection{Vertical Distribution of the Molecular Gas}

\begin{figure}
  \begin{center}
    \FigureFile(110mm,110mm){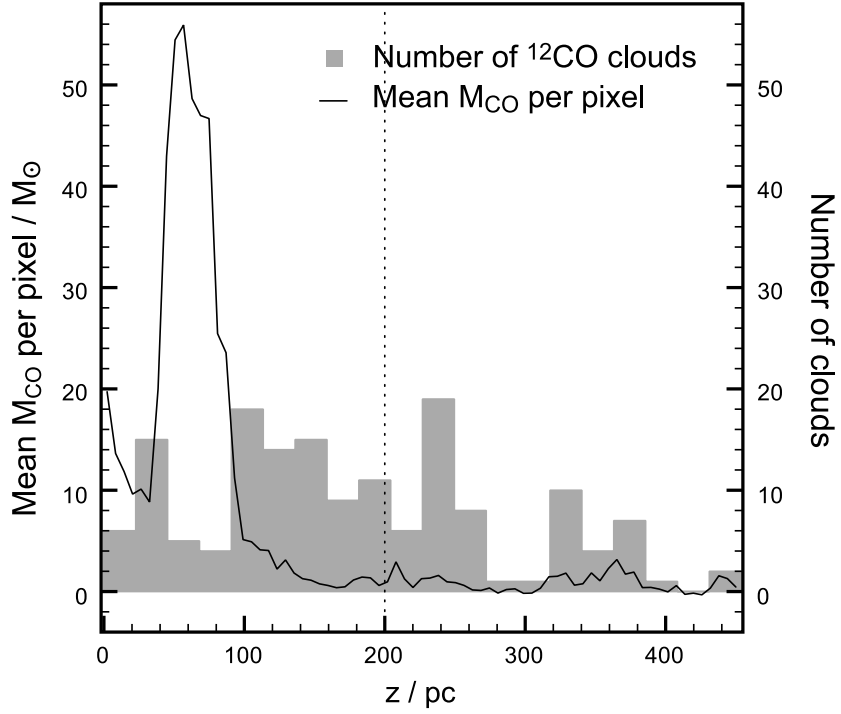}
  \end{center}
  \caption{The vertical distribution of the molecular gas traced in $^{12}$CO(J=1-0), where $z$ refers to distance above the Galactic midplane, assumed to be at $b=0^{\circ}$. The solid line shows the mean value of $M_{\mathrm{CO}}$ per $2'\times2'$ pixel, computed for horizontal bins of $\Delta b = 8'$. The grey histogram shows the number of $^{12}$CO clouds in $\Delta b = 0.5^{\circ}$ bins. The dotted line marks $z=200$ pc, the $3.3\sigma_z$ value for a Gaussian distribution with $\sigma_z = 60$ pc.}
  \label{vmassfig}
\end{figure}

Figure \ref{vmassfig} shows the variation of mean $M_{\mathrm{CO}}$ per $2'\times2'$ pixel with altitude above the midplane, $z$. Also shown is a histogram of cloud count versus $z$. The vertical mass distribution may be discussed quantitatively in terms of a standard Galactic molecular gas scale height. The Carina Flare clouds are located at a Galactic radius of $R\approx8$ kpc, assuming $R_{\solar}=8.5$ kpc \citep{dawson08}. For the molecular disk near the solar circle, the standard value for the Gaussian scale height of the gas is $\sigma_z\approx60$ pc \citep{bronfman88, clemens88, malhotra94a}. For a distribution that follows this model we would therefore expect to find only $\sim0.1\%$ of the total mass at latitudes greater than $3.3\sigma_z$ ($\approx200$ pc). For the present dataset the figure is $\sim7\%$ -- nearly two orders of magnitude greater. Moreover, this fraction includes a large contribution from sub-complexes as high as $\sim6\sigma_z$ and above.\par

Previous studies report some evidence for two different molecular cloud populations in the Galaxy, with differing scale heights, and possibly also with different physical properties \citep{dame94, malhotra94b}. \citet{dame94} find that large-scale CO survey data from the first quadrant is best fitted by two Gaussians rather than one, with a thin layer of $\sigma_z\sim30$ pc and a thick, less massive component  of $\sigma_z\sim100$ pc. Similarly, \citet{malhotra94b} has examined the properties of a distinct population of high-$z$ clouds that are outliers to her single-component gaussian fits. Though the data were less sensitive than the present observations, and the $z$ coverage lower, we find that these outliers share many similarities with our high-$z$ catalogue clouds in terms of size, mass and typical velocity dispersions.\par

Supershells have of course been suggested as one potential means of supporting this high-$z$ population -- either through the transport of existing clouds, or through in-situ formation from swept up atomic gas. The present catalogue provides solid observational proof of high-altitude molecular clouds that are not only robustly associated with a known supershell, but which also share similarities with the previously identified thick-disk component. Although a single object can offer no insight into the relative contribution of supershells to the support of high altitude clouds on a Galaxy-wide scale, the present work does offer a strong indication that they play a role in that support. 

It should be noted that the limited longitude coverage of the catalogue at $b < 2^{\circ}$ results in the selective inclusion of the very mass-rich region of the GMC and the exclusion of the region around it, potentially biasing the results. An alternative treatment might choose to include the full range $284^{\circ} < l < 292^{\circ}$, the overall CO distribution of which is shown in figure 10 of \citet{fukui99}. This would have the effect of increasing the contribution from emission physically distant from the supershell, but would avoid unnaturally inflating the average mass at the latitude of the GMC. We find that the inclusion of these non-catalogue regions does not affect the overall proportion of the mass found at high latitudes, which remains at $\sim7\%$ for $z > 200$ pc. Since the low latitude contribution as a whole is likely to be inflated by the contribution of distant emission, this figure is a lower limit. \\\par

\subsection{Remarks on Mass Estimation}

It is worth making a few comments on the uncertainties associated with molecular cloud mass estimation, which are often passed over in CO cloud studies.

The empirical `X-factor', used to convert $^{12}$CO luminosity to H$_2$ column density, has been derived from a number of independent methods, and is believed to be accurate to a factor of around two, when averaged over large cloud populations \citep[see e.g.][]{dame01, hunter97, solomon91}. For individual clouds the uncertainty will be larger \citep[e.g.][]{magnani95}. In particular, \citet{heyer01} have demonstrated how the correct X-factor for a cloud depends on the gravitational state of the gas, and argue that it may significantly overestimate the masses of objects that are not bound by self-gravity. Conversely, several authors have used the results of PDR modeling to argue that the standard Galactic X-factor may significantly \textit{underestimate} the masses of many molecular clouds \citep{bell06, kaufman99, pringle01}, especially those that are particularly diffuse.

$^{13}$CO LTE-based mass estimates are also subject to their own uncertainties. \citet{padoan00} 
use model clouds with realistic temperature and density structures to argue that LTE analysis may underestimate true column density by a factor of 1.3 to 7, with the most extreme differences occurring in regions of low column density. The LTE values of $N(^{13}\mathrm{CO})$ observed in the present dataset range from $\sim5\times10^{14}$ cm$^{-2}$ to $\sim8\times10^{15}$ cm$^{-2}$, which corresponds to a factor of $N(^{13}\mathrm{CO})_{true}/N(^{13}\mathrm{CO})_{LTE}~\lesssim~3.5$ in their analysis. The unavoidable assumption of a constant H$_2$ to $^{13}$CO abundance ratio may also introduce error into the results.

CO catalogues such as the one presented here offer an attractive means of calibrating molecular cloud mass conversion factors, in particular the X-factor, through comparison with data from the next generation of gamma ray telescopes such as GLAST. Molecular clouds act as passive targets for hadronic cosmic ray interactions, which result in the production of gamma rays in the MeV to TeV regime. The mass of the target cloud may be calculated directly from the gamma-ray luminosity, and this mass may be compared to CO luminosity without the need for many of the assumptions that plague other calibration methods. 

Unfortunately, the estimated masses and angular extents of the catalogue clouds place most of them well below the detectability threshold for GLAST \citep{torres05}. Nevertheless, several high-latitude objects -- notably clouds 74 and 16 -- fall on the border of predicted range and may eventually prove to be detectable. In the near future these clouds, and others like them, may offer the opportunity to calibrate the X-factor for high altitude clouds away from the immediate solar neighborhood.

\section{Summary}
\label{summary}

We have presented a catalogue of $^{12}$CO(J=1-0) and $^{13}$CO(J=1-0) molecular clouds in the spatio-velocity range of the Carina Flare supershell, GSH 287+04-17. Previous work suggests that the majority of the included emission forms part of a complex of molecular clouds that are genuinely associated with the expanding supershell \citep{dawson08}. 

1) Decomposition of the emission into discrete regions results in the identification of 156 $^{12}$CO clouds and 60 $^{13}$CO clouds. The detection limit in column density is estimated as $N(\mathrm{H}_2)\gtrsim2.5\times10^{20}$ cm$^{-2}$ for both $^{12}$CO and $^{13}$CO, which results in completeness limits of $\sim60 M_{\solar}$ for regions observed at $2'$ grid spacing, and $\sim 200 M_{\solar}$ for regions observed at $4'$ spacing. The former contain $\sim85\%$ of the cloud sample by number, and over $\sim90\%$ by mass. 

2) The cloud internal velocity dispersions, $\Delta v_{cld}$, are consistent with other Galactic molecular cloud samples of the same size range. We see no evidence for intrinsically broad components that might indicate shocked molecular gas, with the possible exception of the GMC 288.5+1.5. Previous work suggests this object may be interacting with two Galactic supershells. 

3) The mass spectra of the $^{12}$CO(J=1-0) and $^{13}$CO(J=1-0) clouds are best fitted by power laws of $\gamma = -1.53\pm0.06$ and $\gamma = -1.57\pm0.11$, respectively, where $N(>M) \propto M^{\gamma+1}$. These slopes are consistent with those of other inner and local Galaxy cloud samples.

4) The average ratios of luminosity mass to virial mass are $M_{vir}/M_{\mathrm{CO}}=14$ and $M_{vir}/M_{\mathrm{LTE}}=9$ for the $^{12}$CO and $^{13}$CO populations respectively. The lack of virialized clouds in the sample is notable, but it is is also broadly consistent with the lack of high-mass objects in the present catalogue. The sample as a whole shows no statistically significant differences to typical inner Galaxy populations. 

5) The correlation between size, $R_{cld}$, and velocity dispersion, $\Delta v_{cld}$, is relatively weak for both the $^{12}$CO and $^{13}$CO cloud samples, with Spearman's rank correlation coefficients of 0.15 and 0.40 respectively. The large scatter is consistent with the small dynamic range and lack of large objects in the present sample. The $^{13}$CO cloud sample is best fitted by a power law with $\Delta v_{cld} \propto (R_{cld})^{0.44}$.

6) The present cloud sample are distributed at unusually high altitudes above the Galactic disk. The proportion of mass at $z>200$ pc is almost two orders of magnitude greater than expected for a Gaussian vertical density distribution with $\sigma_z=60$ pc, and these high-$z$ catalogue clouds show a similar range of sizes and masses to the high-altitude molecular disk component observed by \citet{dame94} and, \citet{malhotra94b}. The robust association of such objects with an expanding supershell provides some observational backing for the theory that expanding shells contribute to the support of a high-altitude molecular layer.\\\par

We extend our warmest thanks to all the staff and students of Nagoya University who contributed to the observations utilised in this paper, and to the Las Campanas Observatory staff for their support and hospitality. The NANTEN project was based on a mutual agreement between Nagoya University and the Carnegie Institute of Washington, and its operation was made possible thanks to contributions from many companies and members of the Japanese public. This work is financially supported in part by Grants-in-Aid for Scientific Research (KAKENHI) from the Ministry of Education, Culture, Sports, Science and Technology of Japan (Nos. 15071203 and 18026004) and from the JSPS (Nos. 14102003, 20244014, and 18684003). J. Dawson wishes to acknowledge financial support from a Japanese Government (MEXT) scholarship for postgraduate research studies.

\newpage

\newpage

\newpage

\begin{longtable}{ccccccccccccc}
  \caption{Observed properties of the $^{12}$CO(J=1-0) clouds}
  \label{obsproptable12}
  \hline              
  No. & Name & Grid & \multicolumn{4}{c}{Position} & \multicolumn{6}{c}{Observed Properties} \\ 
  \cline{4-13}
  & & & \textit{l} & \textit{b} & $\alpha(\mathrm{2000}) $ & $\delta(\mathrm{2000}) $ & $T_R^*$ & $W_{\mathrm{CO}}$ & $\Delta v$ & $v_{lsr}$ & $\Delta v (\mathrm{cld})$ & $v_{lsr} (\mathrm{cld})$\\
   & & $(')$& ($^{\circ}$) & ($^{\circ}$) & (h m s) & ($^{\circ}$ $'$ $''$) & (K) & (K km s$^{-1}$) & (km s$^{-1}$) & (km s$^{-1}$) & (km s$^{-1}$) & (km s$^{-1}$)\\
\endfirsthead
  \hline
  No. & Name & Grid & \multicolumn{4}{c}{Position} & \multicolumn{6}{c}{Observed Properties} \\ 
  \cline{4-13}
  & & & \textit{l} & \textit{b} & $\alpha(\mathrm{2000}) $ & $\delta(\mathrm{2000}) $ & $T_R^*$ & $W_{\mathrm{CO}}$ & $\Delta v$ & $v_{lsr}$ & $\Delta v (\mathrm{cld})$ & $v_{lsr} (\mathrm{cld})$\\
  & & $(')$& ($^{\circ}$) & ($^{\circ}$) & (h m s) & ($^{\circ}$ $'$ $''$) & (K) & (K km s$^{-1}$) & (km s$^{-1}$) & (km s$^{-1}$) & (km s$^{-1}$) & (km s$^{-1}$) \\
\endhead
  \hline
\endfoot
  \hline
\endlastfoot
  \hline
1 &  284.23+2.10   & 2 & 284.23 & 2.10 &  10 32 59.8  &  -55 39 47.7  & 1.7 & 5.4 & 2.9 & -7.2 & 2.7 &  -7.4 \\
2 &  284.33+2.30   & 2 & 284.33 & 2.30 &  10 34 19.2  &  -55 32 27.0  & 1.2 & 3.9 & 3.1 & -6.4 & 2.7 &  -6.7 \\
3 &  284.80+2.60   & 2 & 284.80 & 2.60 &  10 38 14.1  &  -55 30 40.9  & 1.3 & 2.3 & 1.6 & -10.6 & 1.5 &  -10.7 \\
4 &  284.83+2.43   & 2 & 284.83 & 2.43 &  10 37 51.6  &  -55 40 23.3  & 3.1 & 7.7 & 2.3 & -9.1 & 2.3 &  -9.5 \\
5 &  284.83+2.13   & 2 & 284.83 & 2.13 &  10 36 48.6  &  -55 56 03.5  & 1.2 & 3.4 & 2.5 & -14.9 & 1.7 &  -14.9 \\
6 &  285.00+2.03   & 2 & 285.00 & 2.03 &  10 37 29.8  &  -56 06 12.7  & 1.5 & 5.3 & 3.4 & -5.6 & 2.2 &  -6.3 \\
7 &  285.17+7.00   & 2 & 285.17 & 7.00 &  10 54 12.2  &  -51 46 55.3  & 1.0 & 6.0 & 3.7 & -10.1 & 3.0 &  -10.0 \\
8 &  285.40+5.47   & 2 & 285.40 & 5.47 &  10 51 05.8  &  -53 15 31.3  & 1.2 & 2.4 & 1.9 & -7.8 & 1.9 &  -8.2 \\
\end{longtable}

\newpage

\begin{longtable}{ccccccccccccc}
  \caption{Observed properties of the $^{13}$CO(J=1-0) clouds}
  \label{obsproptable13}
  \hline              
  No. & Parent $^{12}$CO & Grid & \multicolumn{4}{c}{Position} & \multicolumn{6}{c}{Observed Properties} \\ 
  \cline{4-13}
  & Cloud Name & & \textit{l} & \textit{b} & $\alpha(\mathrm{2000}) $ & $\delta(\mathrm{2000}) $ & $T_R^*$ & $W_{\mathrm{13CO}}$ & $\Delta v$ & $v_{lsr}$ & $\Delta v_{cld}$ & $v_{lsr, cld}$\\
   & & $(')$& ($^{\circ}$) & ($^{\circ}$) & (h m s) & ($^{\circ}$ $'$ $''$) & (K) & (K km s$^{-1}$) & (km s$^{-1}$) & (km s$^{-1}$) & (km s$^{-1}$) & (km s$^{-1}$)\\
\endfirsthead
  \hline
  No. & Parent $^{12}$CO & Grid & \multicolumn{4}{c}{Position} & \multicolumn{6}{c}{Observed Properties} \\ 
  \cline{4-13}
  & Cloud Name & & \textit{l} & \textit{b} & $\alpha(\mathrm{2000}) $ & $\delta(\mathrm{2000}) $ & $T_R^*$ & $W_{\mathrm{13CO}}$ & $\Delta v$ & $v_{lsr}$ & $\Delta v_{cld}$ & $v_{lsr, cld}$\\
  & & $(')$& ($^{\circ}$) & ($^{\circ}$) & (h m s) & ($^{\circ}$ $'$ $''$) & (K) & (K km s$^{-1}$) & (km s$^{-1}$) & (km s$^{-1}$) & (km s$^{-1}$) & (km s$^{-1}$) \\
\endhead
  \hline
\endfoot
  \hline
\endlastfoot
  \hline
  15 & 285.67+4.17  & 2  & 285.67  & 4.17  & 10 48 43.3 & -54 32 19.8 & 1.7  & 2.3  & 1.3  & -25.0  & 1.3  & -25.0  \\
16a & 285.90+4.53  & 2  & 285.87  & 5.00  & 10 52 30.0 & -53 53 01.9 & 0.8  & 1.3  & 1.4  & -23.6  & 1.3  & -23.7  \\
16b & ... & 2  & 285.87  & 4.80  & 10 51 53.7 & -54 03 47.0 & 0.8  & 1.5  & 1.7  & -24.0  & 1.6  & -24.0  \\
16c & ... & 2  & 285.87  & 4.40  & 10 50 40.3 & -54 25 15.1 & 1.2  & 1.5  & 1.1  & -25.4  & 1.1  & -25.4  \\
16d & ... & 2  & 285.90  & 4.53  & 10 51 17.2 & -54 18 59.6 & 2.7  & 6.9  & 2.4  & -23.7  & 2.6  & -23.6  \\
16e & ... & 2  & 286.00  & 5.23  & 10 54 00.2 & -53 43 58.5 & 1.6  & 2.0  & 1.2  & -24.5  & 1.5  & -24.2  \\
16f & ... & 2  & 286.00  & 5.07  & 10 53 30.5 & -53 52 57.6 & 0.9  & 0.9  & 0.9  & -24.3  & 0.9  & -24.2  \\
19 & 286.17+3.33  & 2  & 286.17  & 3.33  & 10 49 10.8 & -55 30 28.3 & 0.6  & 1.0  & 1.5  & -11.0  & 1.5  & -10.8  \\
\end{longtable}

\newpage

\begin{longtable}{cccccccccc}
  \caption{Derived properties of the $^{12}$CO(J=1-0) clouds}
  \label{physproptable12}
  \hline              
  No. & Name & \multicolumn{4}{c}{Position} & \multicolumn{4}{c}{Dervied Properties} \\ 
  \cline{3-10}
  & & \textit{l} & \textit{b} & $\alpha(\mathrm{2000}) $ & $\delta(\mathrm{2000}) $ & $A_{cld}$ & $L_{\mathrm{CO}}$ & $M_{\mathrm{CO}}$ & $M_{vir}$\\
   & & ($^{\circ}$) & ($^{\circ}$) & (h m s) & ($^{\circ}$ $'$ $''$) & (pc$^2$) & (pc$^2$ K km s$^{-1}$) & ($M_{\solar}$) & ($M_{\solar}$)\\
\endfirsthead
  \hline
 No. & Name & \multicolumn{4}{c}{Position} & \multicolumn{4}{c}{Dervied Properties} \\ 
  \cline{3-10}
  & & \textit{l} & \textit{b} & $\alpha(\mathrm{2000}) $ & $\delta(\mathrm{2000}) $ & $A_{cld}$ & $L_{\mathrm{CO}}$ & $M_{\mathrm{CO}}$ & $M_{vir}$\\
  & & ($^{\circ}$) & ($^{\circ}$) & (h m s) & ($^{\circ}$ $'$ $''$) & (pc$^2$) & (pc$^2$ K km s$^{-1}$) & ($M_{\solar}$) & ($M_{\solar}$)\\
\endhead
  \hline
\endfoot
  \hline
\endlastfoot
  \hline
1 & 284.23+2.10  & 284.23  & 2.10  & 10 32 59.8 & -55 39 47.7 & 10.7  & 57  & 199  & 2570  \\
2 & 284.33+2.30  & 284.33  & 2.30  & 10 34 19.2 & -55 32 27.0 & 1.5  & 13  & 46  & 991  \\
3 & 284.80+2.60  & 284.80  & 2.60  & 10 38 14.1 & -55 30 40.9 & 3.8  & 15  & 52  & 441  \\
4 & 284.83+2.43  & 284.83  & 2.43  & 10 37 51.6 & -55 40 23.3 & 33.6  & 138  & 481  & 3310  \\
5 & 284.83+2.13  & 284.83  & 2.13  & 10 36 48.6 & -55 56 03.5 & 3.8  & 21  & 74  & 628  \\
6 & 285.00+2.03  & 285.00  & 2.03  & 10 37 29.8 & -56 06 12.7 & 26.7  & 88  & 308  & 2750  \\
7 & 285.17+7.00  & 285.17  & 7.00  & 10 54 12.2 & -51 46 55.3 & 159.4  & 479  & 1670  & 11900  \\
8 & 285.40+5.47  & 285.40  & 5.47  & 10 51 05.8 & -53 15 31.3 & 6.1  & 17  & 59  & 937  \\
\end{longtable}

\newpage

\begin{longtable}{ccccccccccc}
  \caption{Derived properties of the $^{13}$CO(J=1-0) clouds}
  \label{physproptable13}
  \hline              
  No. & Parent $^{12}$CO & \multicolumn{4}{c}{Position} & \multicolumn{5}{c}{Dervied Properties} \\ 
  \cline{3-11}
  & cloud name & \textit{l} & \textit{b} & $\alpha(\mathrm{2000}) $ & $\delta(\mathrm{2000}) $ & $A_{cld}$ & $\tau$ & $N(\mathrm{H}_2)$ & $M_{\mathrm{LTE}}$ &$M_{vir}$\\
   & & ($^{\circ}$) & ($^{\circ}$) & (h m s) & ($^{\circ}$ $'$ $''$) & (pc$^2$) & & (cm$^{-2}$) & ($M_{\solar}$) & ($M_{\solar}$)\\
\endfirsthead
  \hline
 No. & Parent $^{12}$CO & \multicolumn{4}{c}{Position} & \multicolumn{5}{c}{Dervied Properties} \\ 
  \cline{3-11}
  & cloud name & \textit{l} & \textit{b} & $\alpha(\mathrm{2000}) $ & $\delta(\mathrm{2000}) $ & $A_{cld}$ & $\tau$ & $N(\mathrm{H}_2)$ & $M_{\mathrm{LTE}}$ &$M_{vir}$\\
   & & ($^{\circ}$) & ($^{\circ}$) & (h m s) & ($^{\circ}$ $'$ $''$) & (pc$^2$) & & (cm$^{-2}$) & ($M_{\solar}$) & ($M_{\solar}$)\\
\endhead
  \hline
\endfoot
  \hline
\endlastfoot
  \hline
 15 & 285.67+4.17  & 285.67  & 4.17  & 10 48 43.3 & -54 32 19.8 & 21.9  & 0.15  & 1.2  & 269  & 822 \\
16a & 285.90+4.53  & 285.87  & 5.00  & 10 52 30.0 & -53 53 01.9 & 3.6  & 0.07  & 0.6  & 67  & 317 \\
16b & ... & 285.87  & 4.80  & 10 51 53.7 & -54 03 47.0 & 10.5  & 0.07  & 0.8  & 139  & 910 \\
16c & ... & 285.87  & 4.40  & 10 50 40.3 & -54 25 15.1 & 1.3  & 0.11  & 0.8  & 55  & 135 \\
16d & ... & 285.90  & 4.53  & 10 51 17.2 & -54 18 59.6 & 65.4  & 0.27  & 3.9  & 1660  & 5860 \\
16e & ... & 286.00  & 5.23  & 10 54 00.2 & -53 43 58.5 & 42.5  & 0.14  & 1.1  & 496  & 1470 \\
16f & ... & 286.00  & 5.07  & 10 53 30.5 & -53 52 57.6 & 1.3  & 0.08  & 0.4  & 38  & 88 \\
19 & 286.17+3.33  & 286.17  & 3.33  & 10 49 10.8 & -55 30 28.3 & 1.3  & 0.06  & 0.5  & 43  & 261 \\
\end{longtable}

\end{document}